\documentclass[11pt]{article}
\usepackage{epsfig}
\usepackage{amssymb}
\usepackage{amsmath}
\usepackage{amsfonts}

\newtheorem{theorem}{Theorem}
\newtheorem{definition}[theorem]{Definition}
\newtheorem{lemma}[theorem]{Lemma}
\newtheorem{corollary}[theorem]{Corollary}

\newtheorem{conjecture}{Conjecture}

\newcommand{\qed}{$\square$}

\newenvironment{proof}{%
  \noindent{\it Proof.\ }}{%
  \hspace*{\fill}\qed
  \vspace{2ex}}

\newcommand{\calG}{\mathcal{G}}
\newcommand{\calH}{\mathcal{H}}

\newcommand{\calL}{\mathcal{L}}
\newcommand{\calM}{\mathcal{M}}

\newcommand{\calP}{\mathcal{P}}

\newcommand{\calS}{\mathcal{S}}

\newcommand{\calV}{\mathcal{V}}

\newcommand{\bfD}{\mathbf{D}}

\newcommand{\bfM}{\mathbf{M}}

\newcommand{\bfR}{\mathbf{R}}
\newcommand{\bfS}{\mathbf{S}}

\newcommand{\bfV}{\mathbf{V}}

\newcommand{\NP}{{\rm NP}}

\newcommand{\BPP}{{\rm BPP}}
\newcommand{\BQP}{{\rm BQP}}
\newcommand{\QMA}{{\rm QMA}}

\newcommand{\NIQPZK}{{\rm NIQPZK}}

\newcommand{\SZK}{{\rm SZK}}

\newcommand{\NISZK}{{\rm NISZK}}
\newcommand{\NIQSZK}{{\rm NIQSZK}}
\newcommand{\HVSZK}{{\rm HVSZK}}
\newcommand{\HVQSZK}{{\rm HVQSZK}}

\newcommand{\tr}{\mathrm{tr}}

\newcommand{\init}{\psi_{\mathrm{init}}}

\newcommand{\bra}[1]{\langle #1 |}
\newcommand{\ket}[1]{| #1 \rangle}

\newcommand{\ketbra}[1]{| #1 \rangle \langle #1 |}

\newcommand{\trnorm}[1]{\Vert #1 \Vert_{\mathrm{tr}}}

\newcommand{\Natural}{\mathbb{N}}
\newcommand{\integers}{\mathbb{Z}}
\newcommand{\nonnegative}{\integers^{+}}

\begin{document}
\sloppy

\title{\Large\bf
  Non-Interactive Quantum Statistical and Perfect Zero-Knowledge
}

\author{
  {\large\bf Hirotada Kobayashi}\\
  ~\\
  Quantum Computation and Information Project\\
  Exploratory Research for Advanced Technology\\
  Japan Science and Technology Corporation\\
  5-28-3 Hongo, Bunkyo-ku, Tokyo 113-0033, Japan\\
  ~\\
  {\tt
    hirotada@qci.jst.go.jp
  }\\
  ~\\ 
}

\date{27 July 2002}

\maketitle

\begin{abstract}
  This paper introduces quantum analogues of non-interactive
  perfect and statistical zero-knowledge proof systems.
  Similar to the classical cases,
  it is shown that sharing randomness or entanglement is necessary
  for non-trivial protocols of non-interactive quantum
  perfect and statistical zero-knowledge.
  It is also shown that, with sharing EPR pairs a priori,
  the class of languages having one-sided bounded error
  non-interactive quantum perfect zero-knowledge proof systems
  has a natural complete problem.
  Non-triviality of such a proof system is based on the fact
  proved in this paper that
  the Graph Non-Automorphism problem, which is not known in $\BQP$,
  can be reduced to our complete problem.
  Our results may be the first non-trivial quantum zero-knowledge proofs
  secure even against dishonest quantum verifiers,
  since our protocols are non-interactive, and thus
  the zero-knowledge property does not depend on
  whether the verifier in the protocol is honest or not.
  A restricted version of our complete problem
  derives a natural complete problem for $\BQP$.
\end{abstract}

\section{Introduction}
\label{Section: Introduction}

Zero-knowledge proof systems were introduced
by Goldwasser, Micali, and Rackoff~\cite{GolMicRac89SIComp}
and have been studied extensively
from both complexity theoretical and cryptographic viewpoints.
Because of their wide applicability
in the domain of classical communication and cryptography,
quantum analogue of zero-knowledge proof systems is expected
to play very important roles
in the domain of quantum communication and cryptography.

Very recently Watrous~\cite{Wat02FOCS} proposed
a formal model of quantum statistical zero-knowledge proof systems.
To our knowledge, his model is the only one
for a formal model of quantum zero-knowledge proofs,
although he only considers the case with an {\em honest verifier\/}.
The reason why he only considers the case with an honest verifier
seems to be that
even his model may not give a cryptographically satisfying definition
for quantum statistical zero-knowledge
when the honest verifier assumption is absent.
Indeed, generally speaking,
difficulties arise when we try to define
the notion of quantum zero-knowledge against cheating verifiers
by extending classical definitions of zero-knowledge
in the most straightforward ways.
See~\cite{Gra97PhD} for a discussion of such difficulties
in security of quantum protocols.
Nevertheless, the model of quantum statistical zero-knowledge proofs
by Watrous is natural and reasonable
at least in some restricted situations.
One of such restricted situations is the case with an honest verifier,
which was discussed by Watrous himself.
Another situation is the case of {\em non-interactive\/} protocols,
which this paper treats.

Classical version of non-interactive zero-knowledge proof systems
was introduced by Blum, Feldman, and Micali~\cite{BluFelMic88STOC},
and was later studied by
a number of works~\cite{DeSMicPer87CRYPTO, DeSMicPer88CRYPTO, BluDeSMicPer91SIComp, GolOre94JCrypto, KilPet98JCrypto, DeSDiCPerYun98ICALP, GolSahVad99CRYPTO, Vad99PhD}.
Such non-interactive proof systems put an assumption
that a verifier and a prover share some random string,
and it is known that sharing randomness is necessary
for non-trivial protocols (i.e. protocols for languages beyond $\BPP$)
of non-interactive quantum zero-knowledge proofs~\cite{GolOre94JCrypto}.
As for non-interactive statistical zero-knowledge proof systems,
De~Santis, Di~Crescenzo, Persiano, and Yung showed an existence of
a complete promise problem for the class $\NISZK$ of languages
having non-interactive statistical zero-knowledge proof systems.
Goldreich, Sahai, and Vadhan~\cite{GolSahVad99CRYPTO}
showed another two complete promise problems for $\NISZK$,
namely the Entropy Approximation (EA) problem
and the Statistical Difference from Uniform (SDU) problem,
from which they derived a number of properties of $\NISZK$
such as evidence of non-triviality of the class $\NISZK$.

This paper focuses on quantum analogues of
non-interactive perfect and statistical zero-knowledge proof systems.
The notion of quantum zero-knowledge used in this paper
is along the lines defined by Watrous~\cite{Wat02FOCS}.
First, similar to the classical cases,
it is shown that sharing randomness or entanglement is necessary
for non-trivial protocols (i.e. protocols for languages beyond $\BQP$)
of non-interactive quantum perfect and statistical zero-knowledge.
Next, it is shown that, with sharing EPR pairs a priori,
the class of languages having one-sided bounded error
non-interactive quantum perfect zero-knowledge proof systems
has a natural complete promise problem,
which we call 
the {\em Quantum State Closeness to Identity (QSCI)\/} problem,
informally described as follows:
given a description of a quantum circuit $Q$,
is the output qubits of $Q$ is maximally
entangled to the non-output part or is it far from that?
Note that our QSCI problem may be viewed as a quantum variant of
the SDU problem,
which is shown $\NISZK$-complete
by Goldreich, Sahai, and Vadhan~\cite{GolSahVad99CRYPTO}.
However, our proof for the completeness of the QSCI problem
is quite different from their proof for the classical case
at least in the following two senses:
(i) the completeness of the QSCI problem is shown in a direct manner,
while that of the classical SDU problem was shown
by using other complete problems such as the EA problem,
and (ii) our proof for the completeness result is rather quantum information theoretical.
Using our complete problem,
it is straightforward to show that
the Graph Non-Automorphism (GNA) problem
(or sometimes called the Rigid Graphs problem)
has a non-interactive quantum perfect zero-knowledge proof system
of perfect completeness.
Since the GNA problem is not know in $\BQP$,
this gives an evidence of non-triviality of our proof systems.
One of the merits of considering non-interactive models
is that the zero-knowledge property in non-interactive protocols
does not depend on whether the verifier in the protocol is honest or not.
Thus, our results may be the first non-trivial quantum zero-knowledge proofs
secure even against dishonest quantum verifiers.
It is also shown that a restricted version of our complete problem
derives a natural complete problem for $\BQP$.

The remainder of this paper is organized as follows.
In Section~\ref{Section: Definitions} we give formal definitions
of non-interactive quantum statistical and perfect zero-knowledge proof systems,
and introduce the Quantum State Closeness to Identity problem.
In Section~\ref{Section: Necessity of Shared Entanglement}
we show the necessity of sharing randomness or entanglement
for non-trivial protocols of non-interactive quantum zero-knowledge.
In Section~\ref{Section: NIQPZK-completeness}
we show our main result of completeness and its applications.
Finally, we conclude with Section~\ref{Section: Conjectures},
which mentions our conjectures on non-interactive quantum zero-knowledge proofs.
Familiarity with the basics of quantum computation and information theory
as well as classical zero-knowledge proof systems is assumed throughout this paper.
See~\cite{Gru99Book, NieChu00Book} for the basics of quantum computation and information theory
and~\cite{Gol99Book, Gol01Book} for those of classical zero-knowledge proof systems.

\section{Definitions}
\label{Section: Definitions}

\subsection{Quantum Circuits and Polynomial-Time Preparable Sets of Quantum States}

A family $\{Q_{x} \}$ of quantum circuits is said to be
{\em polynomial-time uniformly generated\/}
if there exists a classical deterministic procedure
that, on each input $x$, outputs a description of $Q_{x}$
and runs in time polynomial in $n = |x|$.
For simplicity,
we assume that
all input strings are over the alphabet $\Sigma = \{ 0, 1 \}$.
It is assumed that the quantum circuits
in such a family are composed of gates
in some reasonable, universal, finite set of quantum gates
such as the Shor basis.
Furthermore, it is assumed that the number of gates in any circuit is not more than
the length of the description of that circuit,
therefore $Q_{x}$ must have size polynomial in $n$.
For convenience, in the subsequent sections,
we often identify a circuit $Q_{x}$ with the unitary operator it induces.

It should be mentioned that
to permit non-unitary quantum circuits, in particular,
to permit measurements at any timing in the computation
does not change the computational power of the model of quantum circuits
in view of time complexity.
See~\cite{AhaKitNis98STOC} for a detailed description
of the equivalence of the unitary and non-unitary quantum circuit models.

Given a collection $\{ \rho_{x} \}$ of mixed states,
let us say that the collection is {\em polynomial-time preparable\/}
if there exists a
polynomial-time uniformly generated family $\{ Q_{x} \}$ of quantum circuits
such that, for every $x$ of length $n$,
(i) $Q_{x}$ is a quantum circuit over $q(n)$ qubits
for some polynomially bounded function
$q \colon \nonnegative \rightarrow \Natural$,
and (ii) for the pure state $Q_{x} \ket{0^{q(n)}}$,
the first $q_{\rm out}(n)$ qubits of it is in the mixed state $\rho_{x}$
when tracing out the rest $q(n) - q_{\rm out}(n)$ qubits,
where $q_{\rm out} \colon \nonnegative \rightarrow \Natural$
is a polynomially bounded function satisfying $q_{\rm out} \leq q$.
In the above description,
the collection of the first $q_{\rm out}(n)$ qubits
may be regarded as an output,
and thus we also say that such a family $\{ Q_{x} \}$ of quantum circuits
is {\em $q$-in $q_\mathrm{out}$-out\/}.

\subsection{Non-Interactive Quantum Statistical Zero-Knowledge with Shared EPR-Pairs}

Here we give a definition of
non-interactive quantum statistical (and perfect) zero-knowledge proof systems
in which the verifier and the prover share EPR-pairs prior to the protocol.

Similar to quantum statistical zero-knowledge proof systems~\cite{Wat02FOCS},
we define non-interactive quantum statistical zero-knowledge proof systems
in terms of quantum circuits.

For each input $x \in \Sigma^{\ast}$ of length $n = |x|$,
the entire system of a non-interactive quantum statistical zero-knowledge proof
consists of
$q(n)
= q_{\calV}(n) + q_{\calM}(n) + q_{\calP}(n)$ qubits,
where $q_{\calV}(n)$ is the number of qubits
that are private to a verifier $V$,
$q_{\calP}(n)$ is the number of qubits
that are private to a prover $P$,
and $q_{\calM}(n)$ is the number of message qubits
sent from $P$ to $V$.
Furthermore, it is assumed that
the verifier $V$ and the prover $P$ shares EPR pairs a priori
among their private qubits.
Let $q_{\calS}(n)$ be the number of the EPR pairs shared by $V$ and $P$.
It is also assumed that $q_{\calV}$, $q_{\calM}$, and $q_{\calS}$
are polynomially bounded functions.
Let $q_{\calV_{\overline{\calS}}} = q_{\calV} - q_{\calS}$
and
$q_{\calP_{\overline{\calS}}} = q_{\calP} - q_{\calS}$.

A {\em $(q_{\calV}, q_{\calM})$-restricted quantum verifier\/} $V$
is a polynomial-time computable mapping of the form
$V \colon \Sigma^{\ast} \rightarrow \Sigma^{\ast}$,
where $\Sigma = \{ 0, 1 \}$ is the alphabet set.
$V$ receives a message of at most $q_{\calM}(n)$ qubits from the prover,
and uses at most $q_{\calV}(n)$ qubits for his private space,
including qubits of shared EPR pairs.
For each input $x \in \Sigma^{\ast}$ of length $n = |x|$,
$V(x)$ is interpreted as a description of
a polynomial-time uniformly generated quantum circuit acting on
$q_{\calV}(n) + q_{\calM}(n)$ qubits.
One of the private qubits of the verifier is designated as the output qubit.

A {\em $(q_{\calM}, q_{\calP})$-restricted quantum prover\/} $P$
is a mapping of the form
$P \colon \Sigma^{\ast} \rightarrow \Sigma^{\ast}$.
$P$ uses at most $q_{\calP}(n)$ qubits for his private space,
including qubits of shared EPR pairs,
and sends a message of at most $q_{\calM}(n)$ qubits to the verifier.
For each input $x \in \Sigma^{\ast}$, $|x| = n$,
$P(x)$ is interpreted as a description of a quantum circuit
acting on $q_{\calM}(n) + q_{\calP}(n)$ qubits.
No restrictions are placed on the complexity of the mapping $P$
(i.e., each $P(x)$ can be an arbitrary unitary transformation).

A {\em $(q_{\calV}, q_{\calM}, q_{\calP})$-restricted
non-interactive quantum proof system\/}
consists of
a $(q_{\calV}, q_{\calM})$-restricted quantum verifier $V$
and a $(q_{\calM}, q_{\calP})$-restricted quantum prover $P$.
Let $\calV = l_{2}(\Sigma^{q_{\calV}})$,
$\calM = l_{2}(\Sigma^{q_{\calM}})$,
and $\calP = l_{2}(\Sigma^{q_{\calP}})$
denote the Hilbert spaces
corresponding to the private qubits of the verifier,
the message qubits between the verifier and the prover,
and the private qubits of the prover, respectively.
We say that a $(q_{\calV}, q_{\calM}, q_{\calP})$-restricted
non-interactive quantum proof system
is {\em $q_{\calS}$-shared-EPR-pairs\/}
if, for every input $x$ of length $n$,
there are $q_{\calS}(n)$ copies of the EPR pair
$(\ket{00} + \ket{11})/\sqrt{2}$ that are initially shared
by the verifier and the prover.
Let $\calV_{\calS} = l_{2}(\Sigma^{q_{\calS}})$
and $\calP_{\calS} = l_{2}(\Sigma^{q_{\calS}})$
denote the Hilbert spaces
corresponding to the verifier and the prover parts of
these shared EPR pairs, respectively,
and write $\calV = \calV_{\overline{\calS}} \otimes \calV_{\calS}$
and $\calP = \calP_{\overline{\calS}} \otimes \calP_{\calS}$.
It is assumed that
all the qubits in
$\calV_{\overline{\calS}}$, $\calM$, and $\calP_{\overline{\calS}}$
are initialized to the $\ket{0}$-states.

Given a verifier $V$, a prover $P$, and an input $x$ of length $n$,
define a circuit $(P(x), V(x))$ acting on $q(n)$ qubits to be the one
applying $P(x)$ to $\calM \otimes \calP$
and $V(x)$ to $\calV \otimes \calM$ in sequence.

The probability
that the $(P, V)$ accepts $x$ is defined
to be the probability that an observation of the output qubit
in the basis of $\{\ket{0}, \ket{1}\}$ yields $\ket{1}$,
after the circuit $(P(x), V(x))$
is applied to the initial state $\ket{\init}$.

In what follows, the circuits $P(x)$ and $V(x)$ of prover and verifier
may be simply denoted by $P$ and $V$, respectively,
if it is not confusing.
We also use the notation $\bfD(\calH)$ for the set of mixed states in $\calH$.

First we define the class
$\NIQSZK(q_{\calV}, q_{\calM}, q_{\calP}, q_{\calS}, a, b)$
of languages having $(q_{\calV}, q_{\calM}, q_{\calP})$-restricted
$q_{\calS}$-shared-EPR-pairs
non-interactive quantum statistical zero-knowledge proof systems
with error probabilities $a$ and $b$
in completeness and soundness sides,
respectively.

\begin{definition}
Given polynomially bounded functions
$q_{\calV}, q_{\calM}, q_{\calS} \colon \nonnegative \rightarrow \Natural$
and a function $q_{\calP} \colon \nonnegative \rightarrow \Natural$,
and functions $a,b \colon \nonnegative \rightarrow [0,1]$,
let $\NIQSZK(q_{\calV}, q_{\calM}, q_{\calP}, q_{\calS}, a, b)$
denote the class of languages $L$
for which there exists
a $(q_{\calV}, q_{\calM})$-restricted quantum verifier $V$
such that,
for every input $x$ of length $n$,
\begin{itemize}
\item[(i)]
Completeness:\\
if $x \in L$,
there exists a $(q_{\calM}, q_{\calP})$-restricted quantum prover $P$
such that $(P, V)$ accepts $x$ with probability at least $a(n)$,
\item[(ii)]
Soundness:\\
if $x \not\in L$,
for any $(q_{\calM}, q_{\calP})$-restricted quantum prover $P'$,
$(P', V)$ accepts $x$ with probability at most $b(n)$,
\item[(iii)]
Zero-Knowledge:\\
there exists a polynomial-time preparable set $\{ \sigma_{x} \}$
of mixed states of $q_{\calV}(n) + q_{\calM}(n)$ qubits
such that, if $x \in L$,
\[
\trnorm{\sigma_{x} - \tr_{\calP} (P \ketbra{\init} P^{\dagger})} \leq \delta(n)
\]
for an honest prover $P$ and some negligible function $\delta$
(i.e., $\delta(n) < 1/p(n)$ for sufficiently large $n$ for all polynomials $p$).
\end{itemize}
\end{definition}

We say that a language $L$ is in $\NIQSZK(a, b)$ in short
if there exist some polynomially bounded functions
$q_{\calV}$, $q_{\calM}$, and $q_{\calS}$
such that $L$ is in
$\NIQSZK(q_{\calV}, q_{\calM}, q_{\calP}, q_{\calS}, a, b)$
for any function $q_{\calP}$.

Similarly, we define the class
$\NIQPZK(q_{\calV}, q_{\calM}, q_{\calP}, q_{\calS}, a, b)$
of languages having $(q_{\calV}, q_{\calM}, q_{\calP})$-restricted
$q_{\calS}$-shared-EPR-pairs
non-interactive quantum perfect zero-knowledge proof systems
with error probabilities $a$ and $b$
in completeness and soundness sides,
respectively.

\begin{definition}
Given polynomially bounded functions
$q_{\calV}, q_{\calM}, q_{\calS} \colon \nonnegative \rightarrow \Natural$
and a function $q_{\calP} \colon \nonnegative \rightarrow \Natural$,
and functions $a,b \colon \nonnegative \rightarrow [0,1]$,
let $\NIQPZK(q_{\calV}, q_{\calM}, q_{\calP}, q_{\calS}, a, b)$
denote the class of languages $L$
for which there exists
a $(q_{\calV}, q_{\calM})$-restricted quantum verifier $V$
such that,
for every input $x$ of length $n$,
\begin{itemize}
\item[(i)]
Completeness:\\
if $x \in L$,
there exists a $(q_{\calM}, q_{\calP})$-restricted quantum prover $P$
such that $(P, V)$ accepts $x$ with probability at least $a(n)$,
\item[(ii)]
Soundness:\\
if $x \not\in L$,
for any $(q_{\calM}, q_{\calP})$-restricted quantum prover $P'$,
$(P', V)$ accepts $x$ with probability at most $b(n)$,
\item[(iii)]
Zero-Knowledge:\\
there exists a polynomial-time preparable set $\{ \sigma_{x} \}$
of mixed states of $q_{\calV}(n) + q_{\calM}(n)$ qubits
such that, if $x \in L$,
$\sigma_{x}$ exactly coincides with
$\tr_{\calP} (P \ketbra{\init} P^{\dagger})$.
\end{itemize}
\end{definition}

As is the statistical zero-knowledge case,
we say that a language $L$ is in $\NIQPZK(a, b)$ in short
if there exist some polynomially bounded functions
$q_{\calV}$, $q_{\calM}$, and $q_{\calS}$
such that $L$ is in
$\NIQPZK(q_{\calV}, q_{\calM}, q_{\calP}, q_{\calS}, a, b)$
for any function $q_{\calP}$.

Note that, similar to the $\QMA$ case,
parallel repetition of
non-interactive quantum statistical (or perfect) zero-knowledge proof systems
can reduce completeness and soundness errors to be exponentially small
while preserving the zero-knowledge property.

\subsection{Variants of Quantum State Distinguishability Problem}

This paper focuses on the following promise problems,
all of which are parameterized by constants $\alpha$ and $\beta$
satisfying $0 \leq \alpha < \beta \leq 1$.

First we review the $(\alpha,\beta)$-Quantum State Distinguishability
($(\alpha,\beta)$-QSD) problem,
which was introduced
and shown to be $\HVQSZK$-complete (for any $0 \leq \alpha < \beta^{2} \leq 1$)
by Watrous~\cite{Wat02FOCS}.
Note that this problem can be regarded as a quantum analogue
of the Statistical Difference problem~\cite{SahVad97FOCS},
which is $\HVSZK$-complete (and thus $\SZK$-complete
from the result $\HVSZK = \SZK$~\cite{GolSahVad98STOC} shown later).

\begin{center}

\vspace{2mm}
\underline{$(\alpha, \beta)$-Quantum State Distinguishability
($(\alpha, \beta)$-QSD)}
\vspace{2mm}

\begin{tabular}{@{}lp{14cm}}
Input:
&
  Descriptions of quantum circuits $Q_0$ and $Q_1$,
  each acting over the Hilbert space
  $\calH_\mathrm{in} = \calH_\mathrm{out} \otimes \calH_{\overline{\mathrm{out}}}$,
  where $\calH_\mathrm{in}$ consists of $q_\mathrm{in}$ qubits
  and $\calH_\mathrm{out}$ consists of $q_\mathrm{out} \leq q_\mathrm{in}$ qubits.
\\[1mm]
Promise:
&
  Letting
  $\rho_i =
     \tr_{\calH_{\overline{\mathrm{out}}}}
       (Q_{i} \ketbra{0^{q_\mathrm{in}}} Q_{i}^{\dagger})$
  for $i=0,1$, we have either one of the following two:
  \begin{itemize}
    \item[(a)]
      $\trnorm{\rho_0 - \rho_1} \leq \alpha$,
    \item[(b)]
      $\trnorm{\rho_0 - \rho_1} \geq \beta$.
  \end{itemize}
\\[-1mm]
Output:
&
  Accept if $\trnorm{\rho_0 - \rho_1} \geq \beta$,
  and reject if $\trnorm{\rho_0 - \rho_1} \leq \alpha$.
\\[2mm]
\end{tabular}

\end{center}

Note that the complement of $(\alpha, \beta)$-QSD,
which we call
{\em $(\alpha,\beta)$-Quantum State Closeness
($(\alpha,\beta)$-QSC)\/} problem,
is also $\HVQSZK$-complete, as shown by Watrous~\cite{Wat02FOCS}.

Next we introduce
{\em $(\alpha,\beta)$-Quantum State Closeness to Identity
($(\alpha,\beta)$-QSCI)\/} problem,
which is a restricted version of the $(\alpha,\beta)$-QSC problem.
Later $(0,\beta)$-QSCI problem will be shown to be $\NIQPZK(1,b)$-complete
for any $0 < \beta < 1$ and any bounded error probability $b$.
Note that this problem can be regarded as a quantum analogue
of the Statistical Difference from Uniform Distribution
(SDU) problem~\cite{GolSahVad99CRYPTO},
which is $\NISZK$-complete.

\begin{center}

\vspace{2mm}
\underline{$(\alpha, \beta)$-Quantum State Closeness to Identity
($(\alpha, \beta)$-QSCI)}
\vspace{2mm}

\begin{tabular}{@{}lp{14cm}}
Input:
&
  A description of a quantum circuit $Q$
  acting over the Hilbert space
  $\calH_\mathrm{in} = \calH_\mathrm{out} \otimes \calH_{\overline{\mathrm{out}}}$,
  where $\calH_\mathrm{in}$ consists of $q_\mathrm{in}$ qubits
  and $\calH_\mathrm{out}$ consists of $q_\mathrm{out} \leq q_\mathrm{in}$ qubits.
\\[1mm]
Promise:
&
  Letting
  $\rho =
     \tr_{\calH_{\overline{\mathrm{out}}}}
       (Q \ketbra{0^{q_\mathrm{in}}} Q^{\dagger})$,
  we have either one of the following two:
  \begin{itemize}
    \item[(a)]
      $\trnorm{\rho - I/2^{q_\mathrm{out}}} \leq \alpha$,
    \item[(b)]
      $\trnorm{\rho - I/2^{q_\mathrm{out}}} \geq \beta$.
  \end{itemize}
\\[-1mm]
Output:
&
  Accept if $\trnorm{\rho - I/2^{q_\mathrm{out}}} \leq \alpha$,
  and reject if $\trnorm{\rho - I/2^{q_\mathrm{out}}} \geq \beta$.
\\[2mm]
\end{tabular}

\end{center}

Putting restrictions on the number of output qubits
of the quantum circuits given as input
yields the following two promise problems,
both of which will be shown to be $\BQP$-complete. 

\begin{center}

\vspace{2mm}
\underline{$(\alpha, \beta)$-One Qubit Quantum State Distinguishability
($(\alpha, \beta)$-$1$QSD)}
\vspace{2mm}

\begin{tabular}{@{}lp{14cm}}
Input:
&
  Descriptions of quantum circuits $Q_0$ and $Q_1$,
  each acting over the Hilbert space
  $\calH_\mathrm{in} = \calH_\mathrm{out} \otimes \calH_{\overline{\mathrm{out}}}$,
  where $\calH_\mathrm{in}$ consists of $q_\mathrm{in}$ qubits
  and $\calH_\mathrm{out}$ consists of a single qubit.
\\[1mm]
Promise:
&
  Letting
  $\rho_i =
     \tr_{\calH_{\overline{\mathrm{out}}}}
       (Q_{i} \ketbra{0^{q_\mathrm{in}}} Q_{i}^{\dagger})$
  for $i=0,1$, we have either one of the following two:
  \begin{itemize}
    \item[(a)]
      $\trnorm{\rho_0 - \rho_1} \leq \alpha$,
    \item[(b)]
      $\trnorm{\rho_0 - \rho_1} \geq \beta$.
  \end{itemize}
\\[-1mm]
Output:
&
  Accept if $\trnorm{\rho_0 - \rho_1} \geq \beta$,
  and reject if $\trnorm{\rho_0 - \rho_1} \leq \alpha$.
\\[2mm]
\end{tabular}

\end{center}

\begin{center}

\vspace{2mm}
\underline{$(\alpha, \beta)$-One Qubit Quantum State Closeness to Identity
($(\alpha, \beta)$-$1$QSCI)}
\vspace{2mm}

\begin{tabular}{@{}lp{14cm}}
Input:
&
  A description of a quantum circuit $Q$
  acting over the Hilbert space
  $\calH_\mathrm{in} = \calH_\mathrm{out} \otimes \calH_{\overline{\mathrm{out}}}$,
  where $\calH_\mathrm{in}$ consists of $q_\mathrm{in}$ qubits
  and $\calH_\mathrm{out}$ consists of a single qubit.
\\[1mm]
Promise:
&
  Letting
  $\rho =
     \tr_{\calH_{\overline{\mathrm{out}}}}
       (Q \ketbra{0^{q_\mathrm{in}}} Q^{\dagger})$,
  we have either one of the following two:
  \begin{itemize}
    \item[(a)]
      $\trnorm{\rho - I/2} \leq \alpha$,
    \item[(b)]
      $\trnorm{\rho - I/2} \geq \beta$.
  \end{itemize}
\\[-1mm]
Output:
&
  Accept if $\trnorm{\rho - I/2} \leq \alpha$,
  and reject if $\trnorm{\rho - I/2} \geq \beta$.
\\[2mm]
\end{tabular}

\end{center}

\section{Necessity of Shared Randomness or Shared Entanglement}
\label{Section: Necessity of Shared Entanglement}

First, similar to the classical cases~\cite{GolOre94JCrypto},
it is shown that sharing randomness or entanglement is necessary
for non-trivial protocols of non-interactive quantum
perfect and statistical zero-knowledge.

\begin{theorem}
Without shared randomness nor shared entanglement,
any language having non-interactive quantum
perfect or statistical zero-knowledge proofs is necessarily in $\BQP$.
\end{theorem}

\begin{proof}
It is sufficient to show that,
without shared randomness nor shared entanglement,
$\NIQSZK(3/4, 1/4)$ is in $\BQP$.

Let $L$ be a language having an $\NIQSZK(3/4, 1/4)$ protocol without shared randomness nor shared entanglement.
Let $V$ be the corresponding honest quantum verifier
and $\{ \sigma_{x} \}$ be the corresponding polynomial-time preparable set
of mixed states.
For every input $x$,
consider the following polynomial-time quantum algorithm:

\begin{enumerate}
\item
Prepare $\sigma_{x}$ in a quantum register $\bfR$.
\item
Apply $V(x)$ to $\bfR$ to have a state $V(x)\sigma_{x}V(x)^{\dagger}$.
\item
Accept iff the contents of $\bfR$ correspond to ones
that make the original verifier $V$ accept.
\end{enumerate}

\begin{itemize}
\item[(i)]
In the case $x$ is in $L$:\\
From the zero-knowledge property of the original protocol,
the difference between $\sigma_{x}$
and the state received from the honest prover is negligible.
Thus, the input $x$ is accepted by the algorithm above
with probability more than $2/3$.
\item[(ii)]
In the case $x$ is not in $L$:\\
From the soundness property of the original protocol,
whatever state the honest verifier $V$ receives from the prover,
$V$ accepts the input $x$ with probability at most $1/4$.
In particular, if the honest verifier $V$ receives $\sigma_{x}$
from the prover, 
$V$ accepts the input $x$ with probability at most $1/4$.
Thus, the input $x$ is accepted by the algorithm above
with probability at most $1/4$ (less than $1/3$).
\end{itemize}

\end{proof}

\section{Completeness Results and their Applications}
\label{Section: NIQPZK-completeness}

\subsection{$\boldsymbol{\NIQPZK(1, 1/2)}$-Completeness of $\boldsymbol{(0, \beta)}$-QSCI}

Here we show that the $(0, \beta)$-QSCI problem
is $\NIQPZK(1, 1/2)$-complete, that is, complete for the class
of languages having
non-interactive quantum perfect zero-knowledge proof systems
of perfect completeness.
While our result is closely related to the classical result
by Goldreich, Sahai, and Vadhan~\cite{GolSahVad99CRYPTO},
the proofs adopted in this paper are rather quantum information theoretical.

The proof of Lemma~\ref{Lemma: (0, beta)-QSCI is in NIQPZK(1,1/2)}
below uses the following well-known property in quantum information theory.

\begin{theorem}[\cite{Uhl86RMathPhy, HugJozWoo93PLA}]
Let $\ket{\phi}, \ket{\psi} \in \calH_{1} \otimes \calH_{2}$ satisfy
$\tr_{\calH_{2}} \ket{\phi}\bra{\phi} = \tr_{\calH_{2}} \ket{\psi}\bra{\psi}$.
Then there is a unitary transformation $U$ over $\calH_{2}$
such that $(I_{\calH_{1}} \otimes U) \ket{\phi} = \ket{\psi}$,
where $I_{\calH_{1}}$ is the identity operator over $\calH_{1}$.
\label{Theorem: locality theorem}
\end{theorem}

\begin{lemma}
$(0, \beta)$-QSCI is in $\NIQPZK(1, 1/2)$
for any $0 < \beta < 1$.
\label{Lemma: (0, beta)-QSCI is in NIQPZK(1,1/2)}
\end{lemma}
 
\begin{proof}
Let $Q$ be a quantum circuit of the $(0, \beta)$-QSCI,
which is $q$-in $q_\mathrm{out}$-out.
Running $O(n)$ copies of $Q$ in parallel
for $n$ exceeding the length of the input $Q$
constructs a quantum circuit $R$ of $q'$-in $q'_\mathrm{out}$-out
that outputs the associated mixed state $\xi$ of $q'_\mathrm{out}$ qubits
and $\xi$ is either $I/2^{q'_\mathrm{out}}$ or
the one such that $\trnorm{\xi - I/2^{q_\mathrm{out}}}$ is arbitrary close to $1$,
say $\trnorm{\xi - I/2^{q_\mathrm{out}}} > 1 - 2^{-n}$.

We construct a $(q'_\mathrm{out}, q'-q'_\mathrm{out}, q_{\calP})$-restricted
non-interactive quantum perfect zero-knowledge proof system
of $q'_\mathrm{out}$-shared-EPR-pairs.
Consider the $(q'_\mathrm{out}, q'-q'_\mathrm{out})$-restricted
quantum verifier $V$.
Let the quantum registers $\bfM$ and $\bfS$
consist of the message qubits
and qubits in the verifier part of the shared EPR pairs, respectively. 
The verification procedure of the verifier is as follows:

\begin{enumerate}
\item
Receive a message in $\mathbf{M}$ from the prover.
\item
Apply $R^{\dagger}$ on the pair of quantum registers $(\bfM, \bfS)$.
\item
Accept if $(\bfM, \bfS)$ contains $0^{q'}$, otherwise reject.
\end{enumerate}

For the completeness, suppose that
$\xi = I/2^{q'_\mathrm{out}}$.
Note that the pure state
$\ket{\phi} = (R \ket{0^{q'}}) \otimes \ket{0^{q_{\calP}}}$
of $q' + q_{\calP}$ qubits
is a purification of $\xi$.
Since the initial state $\ket{\init} \in \calV \otimes \calM \otimes \calP$
of $q' + q_{\calP}$ qubits
is a purification of $I/2^{q'_\mathrm{out}}$
and $\xi = I/2^{q'_\mathrm{out}}$,
from Theorem~\ref{Theorem: locality theorem},
there exists a unitary transformation $P$ over $\calM \otimes \calP$
such that
\[
(I \otimes P) \ket{\init} = \ket{\phi}.
\]
Therefore,
\[
(R^{\dagger} \otimes I) (I \otimes P) \ket{\init} = \ket{0^{q' + q_{\calP}}}.
\]
Thus $V$ accepts the input with certainty.

For the soundness, suppose that
$\trnorm{\xi - I/2^{q'_\mathrm{out}}} > 1 - 2^{-n}$.
Then, for any unitary transformation $P'$ over $\calM \otimes \calP$,
letting $\ket{\psi} = (I \otimes P') \ket{\init}$,
we have
\[
\trnorm{\tr_{\calP} \ketbra{\phi} - \tr_{\calP} \ketbra{\psi}} > 1 - 2^{-n},
\]
since $\tr_{\calM} (\tr_{\calP} \ketbra{\phi}) = \xi$
and $\tr_{\calM} (\tr_{\calP} \ketbra{\psi}) = I/2^{q'_\mathrm{out}}$.
Therefore we have,
\[
\trnorm{\ketbra{0^{q'}}
        - R^{\dagger} (\tr_{\calP} \ketbra{\psi}) R}
> 1 - 2^{-n}.
\]
Thus the probability that $V$ accepts the input is negligible.

Finally, the zero-knowledge property is obvious,
because
$R \ketbra{0^{q'}} R^{\dagger}
=\tr_{\calP} ((I \otimes P) \ketbra{\init} (I \otimes P^{\dagger}))$
is polynomial-time preparable.
\end{proof}

\begin{lemma}
For any promise problem $L \in \NIQPZK(1,1/2)$,
there is a polynomial-time deterministic procedure
that reduces $L$ to the $(0,\beta)$-QSCI problem
for $0 < \beta < 1$.
\label{Lemma: NIQPZK(1,1/2)-hardness of (0, beta)-QSCI}
\end{lemma}

\begin{proof}
Let $L$ be in $\NIQPZK(1,1/2)$.
Then from the fact that parallel repetition works well
for non-interactive quantum perfect zero-knowledge proof systems,
for any function $q_{\calP} \colon \nonnegative \rightarrow \Natural$,
$L$ has a $(q_{\calV}, q_{\calM}, q_{\calP})$-restricted
$q_{\calS}$-shared-EPR-pairs
non-interactive quantum perfect zero-knowledge proof system
of perfect completeness 
for some polynomially bounded functions
$q_{\calV}, q_{\calM}, q_{\calS} \colon \nonnegative \rightarrow \Natural$,
whose soundness error is smaller than $2^{-n}$
for inputs of length $n$.

Let $V$ and $P$ be the honest verifier and the honest prover
of this proof system,
and let $V(x)$ and $P(x)$ be the unitary transformations of $V$ and $P$,
respectively, on a given input $x$.
Let $\{ \sigma_{x} \}$ be a polynomial-time preparable set
such that, if the input $x$ of length $n$ is in $L$,
\[
\sigma_{x} = \tr_{\calP} (P(x) \ketbra{\init} P^{\dagger}(x))
\]
for the honest prover $P$.
The existence of such a polynomial-time preparable set is ensured by
the perfect zero-knowledge property.
For convenience, we assume that,
for every input $x$ of length $n$,
the first $q_{\calM}(n)$ qubits of $\sigma_{x}$ correspond to
the message qubits of the original proof system,
the last $q_{\calV}(n) - q_{\calS}(n)$ qubits of $\sigma_{x}$ correspond to
the private qubits of the verifier (not including the prior-entangled part),
and the last qubit corresponds to the output qubit of the original proof system.

Let $\bfM$, $\bfS$, and $\bfV$ be quantum registers,
each of which consists of
$q_{\calM}(n)$ qubits, $q_{\calS}(n)$ qubits,
and $q_{\calV}(n) - q_{\calS}(n)$ qubits respectively.
For every input $x$,
we construct a quantum circuit $Q_x$
that corresponds to the following algorithm:
\begin{enumerate}
\item
Prepare $\sigma_{x}$ in the triplet $(\bfM, \bfS, \bfV)$ of the quantum registers.
\item
If one of qubits in the quantum register $\bfV$ contains $1$,
output $\ketbra{0^{q_{\calS}(n)}}$.
\item
Do one of the following two uniformly at random.
\begin{itemize}
\item[3.1]
Output the qubits in the quantum register $\bfS$.
\item[3.2]
Apply $V(x)$ on the triplet $(\bfM, \bfS, \bfV)$ of the quantum registers.\\
Output $I/2^{q_{\calS}(n)}$ if the last qubit in $\bfV$ contains $1$,
otherwise, output $\ketbra{0^{q_{\calS}(n)}}$.
\end{itemize}
\end{enumerate}

Suppose that $x$ is in $L$.
Then
$
\sigma_{x} = \tr_{\calP} (P(x) \ketbra{\init} P^{\dagger}(x))
$
is satisfied.
Note that
$\tr_{\calV_{\overline{\calS}} \otimes \calM \otimes \calP}
   (P(x) \ketbra{\init} P^{\dagger}(x))
   = I/2^{q_{\calS}(n)}$.
Furthermore, for the state $P(x) \ketbra{\init} P^{\dagger}(x)$,
the verification procedure of $V$ accepts the input with certainty.
Therefore, the circuit $Q_{x}$ constructed above outputs $I/2^{q_{\calS}(n)}$
with certainty.

Now suppose that $x$ is not in $L$.
We claim that the output mixed state $\rho$ of $Q_{x}$ satisfies
$\trnorm{\rho - I/2^{q_{\calS}(n)}} > c$
for some positive constant $c \leq 1$.
Without loss of generality,
we assume that $\sigma_{x}$ is of the form
$\sigma'_{x} \otimes \ketbra{0^{q_{\calV}(n)}}$,
since the step $2$ reduces $\sigma_{x}$ to the state of this form
or outputs the state farthest away from $I/2^{q_{\calS}(n)}$.

For the soundness property of the original proof system,
for any mixed state $\xi \otimes \ketbra{0^{q_{\calV}(n)}}$
in $\bfD(\calM \otimes \calV)$ satisfying
$\tr_{\calM \otimes \calV_{\overline{\calS}}} (\xi \otimes \ketbra{0^{q_{\calV}(n)}})
  = I/2^{q_{\calS}(n)}$,
the verification procedure of $V$ results in accept
with probability at most $2^{-n}$.

Therefore, if
$\trnorm{\tr_{\calM \otimes \calV_{\overline{\calS}}}
           (\sigma'_{x} \otimes \ketbra{0^{q_{\calV}(n)}})
         - I/2^{q_{\calS}(n)}}
 \geq 1/2$,
then
\[
\trnorm{\tr_{\calM} (\sigma'_{x} \otimes \ketbra{0^{q_{\calV}(n)}})
         - I/2^{q_{\calS}(n)} \otimes \ketbra{0^{q_{\calV}(n)}}}
\geq 1/2,
\]
and thus
\[
\trnorm{\tr_{\calM} \sigma'_{x} - I/2^{q_{\calS}(n)}}
\geq 1/2.
\]
Hence the step 3.1 outputs the mixed state $\rho$ satisfying
$\trnorm{\rho - I/2^{q_{\calS}(n)}} \geq 1/2$.

On the other hand, if
$\trnorm{\tr_{\calM \otimes \calV_{\overline{\calS}}}
           (\sigma'_{x} \otimes \ketbra{0^{q_{\calV}(n)}})
         - I/2^{q_{\calS}(n)}}
 \leq 1/2$,
we have
\[
\trnorm{\sigma'_{x} \otimes \ketbra{0^{q_{\calV}(n)}}
         - \xi \otimes \ketbra{0^{q_{\calV}(n)}}}
\leq 1/2
\]
for some mixed state $\xi \otimes \ketbra{0^{q_{\calV}(n)}}$
in $\bfD(\calM \otimes \calV)$ satisfying
$\tr_{\calM \otimes \calV_{\overline{\calS}}} (\xi \otimes \ketbra{0^{q_{\calV}(n)}})
  = I/2^{q_{\calS}(n)}$.
Therefore, the step 3.2 results in rejection with probability
at least $1/2 - 2^{-(n+1)}$,
and thus the circuit $Q_{x}$ outputs $\ketbra{0^{q_{\calV}(n)}}$ with probability
at least $1/2 - 2^{-(n+1)}$.

Putting things together,
in the case $x$ is not in $L$,
the circuit $Q_{x}$ outputs the mixed state $\rho$ satisfying
$\trnorm{\rho - I/2^{q_{\calS}(n)}} > c$
for some constant $c$ greater than, say $1/5$.

Now, constructing $r$ copies of $Q_{x}$ to have a circuit
$Q_{x}^{\otimes r}$ for appropriately chosen $r$
reduces $L$ to the $(0, \beta)$-QSCI problem for arbitrary $0 < \beta < 1$.
\end{proof}

Thus we have the following theorem.

\begin{theorem}
$(0, \beta)$-QSCI is complete for $\NIQPZK(1, 1/2)$
for $0 < \beta < 1$.
\end{theorem}

\subsection{$\boldsymbol{\NIQPZK(1, 1/2)}$-Protocol for Graph Non-Automorphism}

The {\em Graph Non-Automorphism (GNA)\/} problem defined below
is a special case of the {\em graph non-isomorphism (GNI)\/} problem,
and is not known in $\BQP$ nor in $\NP$.

\begin{center}

\vspace{2mm}
\underline{Graph Non-Automorphism (GNA)}
\vspace{2mm}

\begin{tabular}{@{}lp{14cm}}
Input:
&
  A description of a graph $G$ of $n$ vertices.
\\[1mm]
Output:
&
  Accept if $\pi(G) \neq G$ for all non-trivial permutations $\pi$
  over $n$ vertices
  and reject otherwise.
\\[2mm]
\end{tabular}

\end{center}

It is easy to show that any instance of GNA
is reduced to an instance of $(0, \beta)$-QSCI,
and thus we have the following corollary.

\begin{corollary}
  GNA has a non-interactive quantum perfect zero-knowledge proof system
  of perfect completeness.
\end{corollary}

\begin{proof}
We assume an appropriate ordering of permutations over $n$ vertices
so that each permutation can be represented with
$q_{\calL}(n) = \lceil \log n! \rceil = O(n \log n)$ qubits.
Let $\pi_{i}$ be the $i$-th permutation according to this ordering
for $0 \leq i \leq n!-1$.

Let $\calP$ be a Hilbert space consisting of $q_{\calL}(n)$ qubits
and $\calG$ be a Hilbert space consisting of $q_{\calG}(n) = O(n^2)$ qubits
(intuitively, $\calP$ is for a representation of a permutation
and $\calG$ is for a representation of a graph).

Given a graph $G$ of $n$ vertices,
consider the following quantum circuit $Q_{G}$ behaving as follows.

\begin{enumerate}
\item
Prepare the following quantum state in $\calP \otimes \calG$:
\[
\frac{1}{\sqrt{2^{q_{\calL}(n)}}}
  \sum_{i=0}^{n!-1} \ket{i}\ket{0, \pi_{i}(G)}
+
\frac{1}{\sqrt{2^{q_{\calL}(n)}}}
  \sum_{i=n!}^{2^{q_{\calL}(n)}-1} \ket{i}\ket{1, i}.
\]
\item
Output the qubits in $\calP$.
\end{enumerate}

If a given graph $G$ has no non-trivial automorphism groups,
every $\pi_{i}(G)$ is different from each other,
and thus the output of $Q_{G}$ is the mixed state $I/2^{q_{\calL}(n)}$.

On the other hand, if a given graph $G$ has a non-trivial automorphism groups,
the contents of qubits in $\calG$ have at most
$2^{q_{\calL}(n)} - n!/2 \leq 3/4 \cdot 2^{q_{\calL}(n)}$ variations,
and the trace-norm between $I/2^{q_{\calL}(n)}$
and the output of $Q_{G}$ is at least $1/4$.

Thus the constructed quantum circuit $Q_{G}$
is an instance of $(0, 1/4)$-QSDI, which completes the proof.
\end{proof}

\subsection{$\boldsymbol{\BQP}$-Completeness Results}

\begin{theorem}
$(\alpha, \beta)$-1QSCI and $(\alpha, \beta)$-1QSD
are complete for $\BQP$ for $0 < \alpha < \beta < 1$.
\end{theorem}

\begin{proof}
  Straightforward and thus omitted.
\end{proof}

\section{Conjectures}
\label{Section: Conjectures}

\begin{conjecture}
\label{Conjecture: polarize}
There is a (deterministic) polynomial-time procedure that,
on an input $(Q, 1^n)$
where $Q$ is a description of a quantum circuit
specifying a mixed state $\rho$ of $q_1$ qubits,
outputs a description of a quantum circuits $R$
(having size polynomial in $n$ and in the size of $Q$)
specifying a mixed state $\xi$ of $q_2$ qubits satisfying the following
(for $\alpha$ and $\beta$ satisfying an appropriate condition such as
$0 < \alpha < 1/q_1 < 1 - 1/q_1 < \beta < 1$).
\begin{eqnarray*}
\trnorm{\rho - I/2^{q_1}} < \alpha
& \Rightarrow & 
\trnorm{\xi - I/2^{q_2}} < 2^{-n},\\
\trnorm{\rho - I/2^{q_1}} > \beta
& \Rightarrow & 
\trnorm{\xi - I/2^{q_2}} > 1 - 2^{-n}.
\end{eqnarray*}
\end{conjecture}

Under the assumption that Conjecture~\ref{Conjecture: polarize} holds,
the following two conjectures can be shown in similar manners
as the proofs of Lemma~\ref{Lemma: (0, beta)-QSCI is in NIQPZK(1,1/2)}
and Lemma~\ref{Lemma: NIQPZK(1,1/2)-hardness of (0, beta)-QSCI}.

\begin{conjecture}
$(\alpha, \beta)$-QSCI is in $\NIQSZK$
for any $\alpha$ and $\beta$ satisfying an appropriate condition such as
$0 < \alpha < 1/q_1 < 1 - 1/q_1 < \beta < 1$,
where $q_1$ is the number of output qubits of the quantum circuit
given as an instance of $(\alpha, \beta)$-QSCI.
\end{conjecture}
 
\begin{conjecture}
For any promise problem $L \in \NIQSZK$,
there is a polynomial-time deterministic procedure
that reduces $L$ to the $(\alpha,\beta)$-QSCI problem
for any $\alpha$ and $\beta$ satisfying an appropriate condition such as
$0 < \alpha < 1/q_1 < 1 - 1/q_1 < \beta < 1$,
where $q_1$ is the number of output qubits of the quantum circuit
given as an instance of $(\alpha, \beta)$-QSCI.
\end{conjecture}

Thus, under the assumption that Conjecture~\ref{Conjecture: polarize} holds,
the following conjecture is provable.

\begin{conjecture}
$(\alpha, \beta)$-QSCI is complete for $\NIQSZK$
for any $\alpha$ and $\beta$ satisfying an appropriate condition such as
$0 < \alpha < 1/q_1 < 1 - 1/q_1 < \beta < 1$,
where $q_1$ is the number of output qubits of the quantum circuit
given as an instance of $(\alpha, \beta)$-QSCI.
\end{conjecture}


\begin{thebibliography}{10}

\bibitem{AhaKitNis98STOC}
Dorit Aharonov, Alexei~Yu. Kitaev, and Noam Nisan.
\newblock Quantum circuits with mixed states.
\newblock In {\em Proceedings of the Thirtieth Annual ACM Symposium on Theory
  of Computing}, pages 20--30, 1998.

\bibitem{BluDeSMicPer91SIComp}
Manuel Blum, Alfredo {De Santis}, Silvio Micali, and Giuseppe Persiano.
\newblock Non-interactive zero-knowledge.
\newblock {\em SIAM Journal on Computing}, 20(6):1084--1118, 1991.

\bibitem{BluFelMic88STOC}
Manuel Blum, Paul Feldman, and Silvio Mical.
\newblock Non-interactive zero-knowledge and its applications (extended
  abstract).
\newblock In {\em Proceedings of the Twentieth Annual ACM Symposium on Theory
  of Computing}, pages 103--112, 1988.

\bibitem{DeSDiCPerYun98ICALP}
Alfredo {De Santis}, Giovanni {Di Crescenzo}, Giuseppe Persiano, and Moti Yung.
\newblock Image density is complete for non-interactive-{SZK} (extended
  abstract).
\newblock In {\em Proceedings of the 25th International Colloquium on Automata,
  Languages and Programming}, volume 1443 of {\em Lecture Notes in Computer
  Science}, pages 784--795, 1998.

\bibitem{DeSMicPer87CRYPTO}
Alfredo {De Santis}, Silvio Micali, and Giuseppe Persiano.
\newblock Non-interactive zero-knowledge proof systems.
\newblock In {\em Advances in Cryptology -- CRYPTO '87, A Conference on the
  Theory and Applications of Cryptographic Techniques}, volume 293 of {\em
  Lecture Notes in Computer Science}, pages 52--72, 1987.

\bibitem{DeSMicPer88CRYPTO}
Alfredo {De Santis}, Silvio Micali, and Giuseppe Persiano.
\newblock Non-interactive zero-knowledge with preprocessing.
\newblock In {\em Advances in Cryptology -- CRYPTO '88, 8th Annual
  International Cryptology Conference}, volume 403 of {\em Lecture Notes in
  Computer Science}, pages 269--282, 1988.

\bibitem{Gol99Book}
Oded Goldreich.
\newblock {\em Modern Cryptography, Probabilistic Proofs and
  Pseudo-randomness}.
\newblock Springer, 1999.

\bibitem{Gol01Book}
Oded Goldreich.
\newblock {\em Foundations of Cryptography -- Basic Tools}.
\newblock Cambridge, 2001.

\bibitem{GolOre94JCrypto}
Oded Goldreich and Yair Oren.
\newblock Definitions and properties of zero-knowledge proof systems.
\newblock {\em Journal of Cryptology}, 7(1):1--32, 1994.

\bibitem{GolSahVad98STOC}
Oded Goldreich, Amit Sahai, and Salil~P. Vadhan.
\newblock Honest-verifier statistical zero-knowledge equals general statistical
  zero-knowledge.
\newblock In {\em Proceedings of the Thirtieth Annual ACM Symposium on Theory
  of Computing}, pages 399--408, 1998.

\bibitem{GolSahVad99CRYPTO}
Oded Goldreich, Amit Sahai, and Salil~P. Vadhan.
\newblock Can statistical zero knowledge be made non-interactive? or on the
  relationship of {$\SZK$} and {$\NISZK$}.
\newblock In {\em Advances in Cryptology -- CRYPTO '99, 19th Annual
  International Cryptology Conference}, volume 1666 of {\em Lecture Notes in
  Computer Science}, pages 467--484, 1999.

\bibitem{GolMicRac89SIComp}
Shafi Goldwasser, Silvio Micali, and Charles Rackoff.
\newblock The knowledge complexity of interactive proof systems.
\newblock {\em SIAM Journal on Computing}, 18(1):186--208, 1989.
\newblock ~\\Preliminary version appeared in {\em Proceedings of the
  Seventeenth Annual ACM Symposium on Theory of Computing}, pages 291--304,
  1985.

\bibitem{Gra97PhD}
Jeroen van~de Graaf.
\newblock {\em Towards a formal definition of security for quantum protocols}.
\newblock PhD thesis, D\'epartement d'Informatique et de Recherche
  Op\'erationnelle, Universit\'e de Montr\'eal, December 1997.

\bibitem{Gru99Book}
Jozef Gruska.
\newblock {\em Quantum Computing}.
\newblock McGraw-Hill, 1999.

\bibitem{HugJozWoo93PLA}
Lane~P. Hughston, Richard Jozsa, and William~K. Wootters.
\newblock A complete classification of quantum ensembles having a given density
  matrix.
\newblock {\em Physics Letters A}, 183:14--18, 1993.

\bibitem{KilPet98JCrypto}
Joe Kilian and Erez Petrank.
\newblock An efficient noninteractive zero-knowledge proof system for {$\NP$}
  with general assumptions.
\newblock {\em Journal of Cryptology}, 11(1):1--27, 1998.

\bibitem{NieChu00Book}
Michael~A. Nielsen and Isaac~L. Chuang.
\newblock {\em Quantum Computation and Quantum Information}.
\newblock Cambridge University Press, 2000.

\bibitem{SahVad97FOCS}
Amit Sahai and Salil~P. Vadhan.
\newblock A complete promise problem for statistical zero-knowledge.
\newblock In {\em Proceedings of the 38th Annual Symposium on Foundations of
  Computer Science}, pages 448--457, 1997.

\bibitem{Uhl86RMathPhy}
Armin Uhlmann.
\newblock Parallel transport and ``quantum holonomy'' along density operators.
\newblock {\em Reports on Mathematical Physics}, 24:229--240, 1986.

\bibitem{Vad99PhD}
Salil~P. Vadhan.
\newblock {\em A Study of Statistical Zero-Knowledge Proofs}.
\newblock PhD thesis, Department of Mathematics, Massachusetts Institute of
  Technology, August 1999.

\bibitem{Wat02FOCS}
John Watrous.
\newblock Limits on the power of quantum statistical zero-knowledge.
\newblock In {\em Proceedings of the 43rd Annual Symposium on Foundations of
  Computer Science}, 2002.
\newblock To appear.

\end{thebibliography}

\end{document}